\documentclass{elsart}
\usepackage[dvips]{graphicx}
\usepackage{amssymb}
\newcommand{\ie}{i.e.}
\journal{Physica A}
\begin{document}
\begin{frontmatter}
\title{Power Law Tails in the Italian Personal Income Distribution}
\author[Roma,S.I.E.C.]{F. Clementi\corauthref{cor}},
\corauth[cor]{Corresponding author. Tel.: +39--06--49--766--843; fax: +39--06--44--61--964.}
\ead{Fabio.Clementi@uniroma1.it}
\author[Ancona,S.I.E.C.]{M. Gallegati}
\ead{gallegati@dea.unian.it}
\address[Roma]{Department of Public Economics, University of Rome `La Sapienza', Via del Castro Laurenziano 9, I--00161 Rome, Italy}
\address[Ancona]{Department of Economics, Universit\`a Politecnica delle Marche, Piazzale Martelli 8, I--62100 Ancona, Italy}
\address[S.I.E.C.]{S.I.E.C., Universit\`a Politecnica delle Marche, Piazzale Martelli 8, I--62100 Ancona, Italy}
\begin{abstract}
We investigate the shape of the Italian personal income distribution using microdata from the Survey on Household Income and Wealth, made publicly available by the Bank of Italy for the years 1977--2002. We find that the upper tail of the distribution is consistent with a Pareto-power law type distribution, while the rest follows a two-parameter lognormal distribution. The results of our analysis show a shift of the distribution and a change of the indexes specifying it over time. As regards the first issue, we test the hypothesis that the evolution of both gross domestic product and personal income is governed by similar mechanisms, pointing to the existence of correlation between these quantities. The fluctuations of the shape of income distribution are instead quantified by establishing some links with the business cycle phases experienced by the Italian economy over the years covered by our dataset.
\end{abstract}
\end{frontmatter}
\section{Introduction}
In the last decades, extensive literature has shown that the size of a large number of phenomena can be well described by a {\em power law\/} type distribution.
\par
The modeling of income distribution originated more than a century ago with the work of Vilfredo Pareto, who observed in his {\em Cours d'\'economie politique\/} (1897) that a plot of the logarithm of the number of income-receiving units above a certain threshold against the logarithm of the income yields points close to a straight line. This power law behaviour is nowadays known as {\em Pareto law}.
\par
Recent empirical work seems to confirm the validity of Pareto (power) law. For example, \cite{Aoyama_Nagahara_Okazaki_Souma_Takayasu_Takayasu} show that the distribution of income and income tax of individuals in Japan for the year 1998 is very well fitted by a power law, even if it gradually deviates as the income approaches lower ranges. The applicability of Pareto distribution only to high incomes is actually acknowledged; therefore, other kinds of distributions has been proposed by researchers for the low-middle income region. According to \cite{Montroll_Shlesinger}, U.S. personal income data for the years 1935--36 suggest a power law distribution for the high-income range and a lognormal distribution for the rest; a similar shape is found by \cite{Souma} investigating the Japanese income and income tax data for the high-income range over the 112 years 1887--1998, and for the middle-income range over the 44 years 1955--98.\footnote{\cite{Reed} suggests that a Pareto law may hold also for lower incomes, yielding a so-called {\em double Pareto-lognormal distribution}, that is a distribution with a lognormal body and a double Pareto tail.} \cite{Nirei_Souma} confirm the power law decay for top taxpayers in the U.S. and Japan from 1960 to 1999, but find that the middle portion of the income distribution has rather an exponential form; the same is proposed by \cite{Dragulescu_Yakovenko} for the U.K. during the period 1994--99 and for the U.S. in 1998.
\par
The aim of this paper is to look at the shape of the personal income distribution in Italy by using cross-sectional data samples from the population of Italian households during the years 1977--2002. We find that the personal income distribution follows the Pareto law in the high-income range, while the lognormal pattern is more appropriate in the central body of the distribution. From this analysis we get the result that the indexes specifying the distribution change in time; therefore, we try to look for some factors which might be the potential reasons for this behaviour.
\par
The rest of the paper is organized as follows. Sec. \ref{sec:LognormalPatternWithPowerLawTail} reports the data utilized in the analysis and describes the shape of the Italian personal income distribution. Sec. \ref{sec:TimeDevelopmentOfTheDistribution} explains the shift of the distribution and the change of the indexes specifying it over the years covered by our dataset. Sec. \ref{sec:ConcludingRemarks} concludes the paper.
\section{Lognormal pattern with power law tail}
\label{sec:LognormalPatternWithPowerLawTail}
We use microdata from the Historical Archive (HA) of the Survey on Household Income and Wealth (SHIW) made publicly available by the Bank of Italy for the period 1977--2002 \cite{Bank_of_Italy_1}.\footnote{The data for the years preceding 1977 are no longer available. The survey was carried out yearly until 1987 (except for 1985) and every two years thereafter (the survey for 1997 was shifted to 1998). In 1989 a panel section consisting of units already
interviewed in the previous survey was introduced in order to allow for better comparison over time. The basic definition of income provided by the SHIW is net of taxation and social security contributions. It is the sum of
four main components: compensation of employees; pensions and net transfers; net income from self-employment; property income (including income from buildings and income from financial assets). Income from financial assets started to be recorded only in 1987. See \cite{Brandolini} for details on source description, data quality, and main changes in the sample design and income definition.} All amounts are expressed in thousands of lire. Since we are comparing incomes across years, to get rid of inflation data are reported in 1976 prices using the Consumer Prices Index (CPI) issued by the National Institute of Statistics \cite{ISTAT}. The average number of income-earners surveyed from the SHIW-HA is about 10,000.
\par
\begin{figure}[!htbp]
\centering
\includegraphics[width=\textwidth]{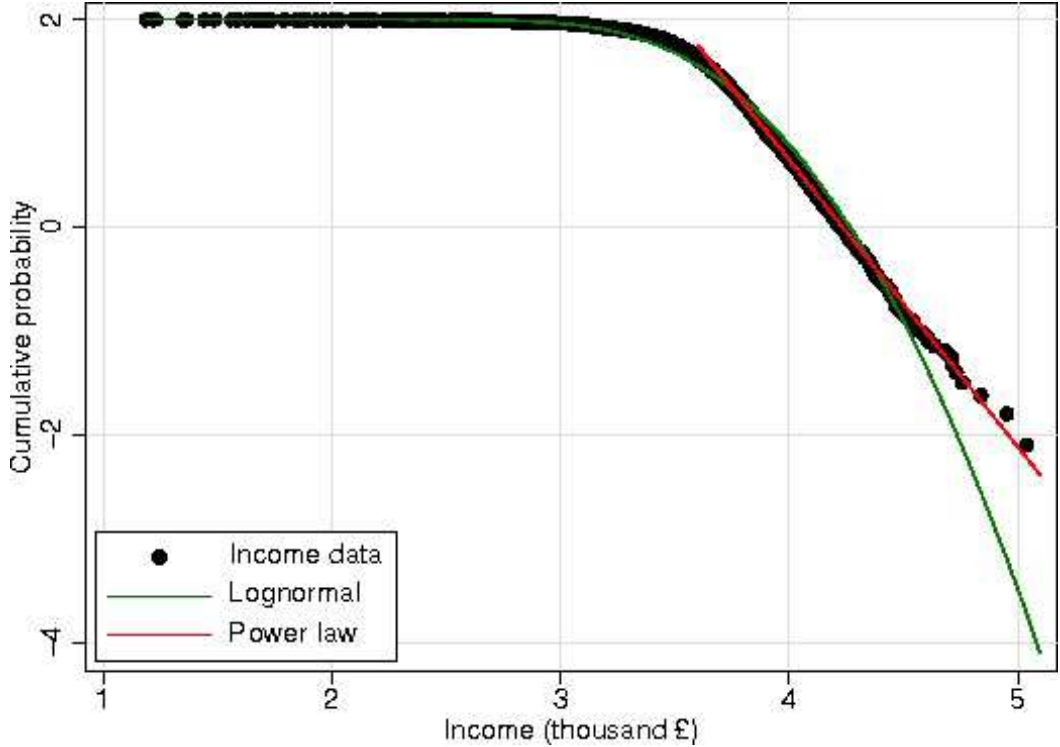}
\caption{The cumulative probability distribution of the Italian personal income in 1998. We take the horizontal axis as the logarithm of the personal income in thousands of lire and the vertical axis as the logarithm of the cumulative probability. The green solid line is the lognormal fit with $\hat{\mu}=3.48$ $(0.004)$ and $\hat{\sigma}=0.34$ $(0.006)$. Gibrat index is $\hat{\beta}=2.10$.}
\label{Cumulative_Probability_1998}
\end{figure}
\begin{figure}[!t]
\centering
\includegraphics[width=\textwidth]{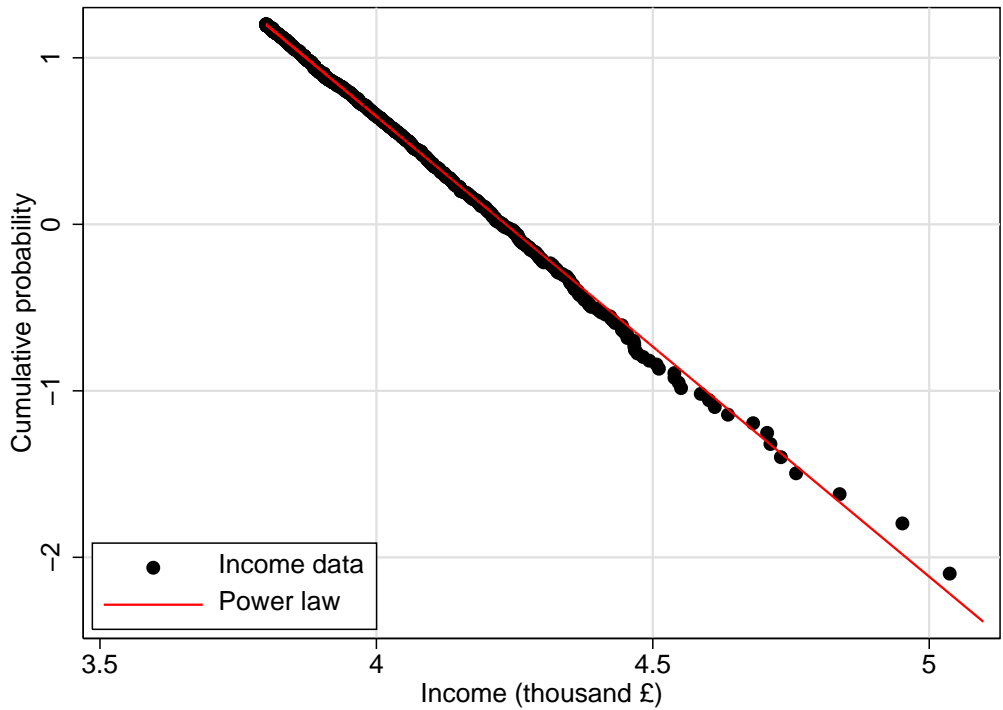}
\caption{The fit to the power law distribution for the year 1998. The red solid line is the best-fit function. Pareto index, obtained by least-square-fit, is $\hat{\alpha}=2.76$ $(0.002)$; the estimated minimum income is $\hat{x}_{0}=17,141$ thousand lire. The goodness of fit of OLS estimate in terms of $R^{2}$ index is 0.9993.}
\label{Power_Law_1998}
\end{figure}

Fig. \ref{Cumulative_Probability_1998}
shows the profile of the personal income distribution for the year 1998. We take the horizontal axis as the logarithm of the income in thousands of lire and the vertical axis as the logarithm of the cumulative probability. The cumulative probability is the probability to find a person with an income greater than or equal to $x$:
\begin{equation}
P(X\geq x)=\int\limits_{x}^{\infty}p(t)dt
\end{equation}
Two facts emerge from this figure. Firstly, the central body of the distribution (almost all of it below the 99\textsuperscript{th} percentile) follows a two-parameter lognormal distribution (green solid line). The probability density function is:
\begin{equation}
p\left(x\right)=\frac{1}{x\sigma\sqrt{2\pi}}exp\left[-\frac{1}{2}\left(\frac{logx-\mu}{\sigma}\right)^{2}\right]
\end{equation}
with $0<x<\infty$, and where $\mu$ and $\sigma$ are the mean and the standard deviation of the normal distribution. The value of the fraction $\beta=\frac{1}{\sqrt{2\sigma^{2}}}$ returns the so-called Gibrat index; if $\beta$ has low values (large variance of the global distribution), the personal income is unevenly distributed. From our dataset we obtain the following maximum-likelihood estimates:\footnote{We exclude from our estimates about the top 1.4\% of the distribution, which behaves as outlier, and about the bottom 0.8\%, which corresponds to non positive entries.} $\hat{\mu}=3.48$ $(0.004)$ and $\hat{\sigma}=0.34$ $(0.006)$;\footnote{The number in parentheses following a point estimate represents its standard error.} Gibrat index is $\hat{\beta}=2.10$. Secondly, about the top 1\% of the distribution follows a Pareto (power law) distribution. This power law behaviour of the tail of the distribution is more evident from Fig. \ref{Power_Law_1998},
where the red solid line is the best-fit linear function. We extract the power law slope (Pareto index) by running a simple OLS regression of the logarithm of the cumulative probability on a constant and the logarithm of personal income, obtaining a point estimate of $\hat{\alpha}=2.76$ $(0.002)$. Given this value for $\hat{\alpha}$, our estimate of $x_{0}$ (the income level below which the Pareto distribution would not apply) is 17,141 thousand lire. The fit of linear regression is extremely good, as one can appreciate by noting that the value of $R^{2}$ index is 0.9993.
\par
\begin{figure}[!t]
\centering
\includegraphics[width=\textwidth]{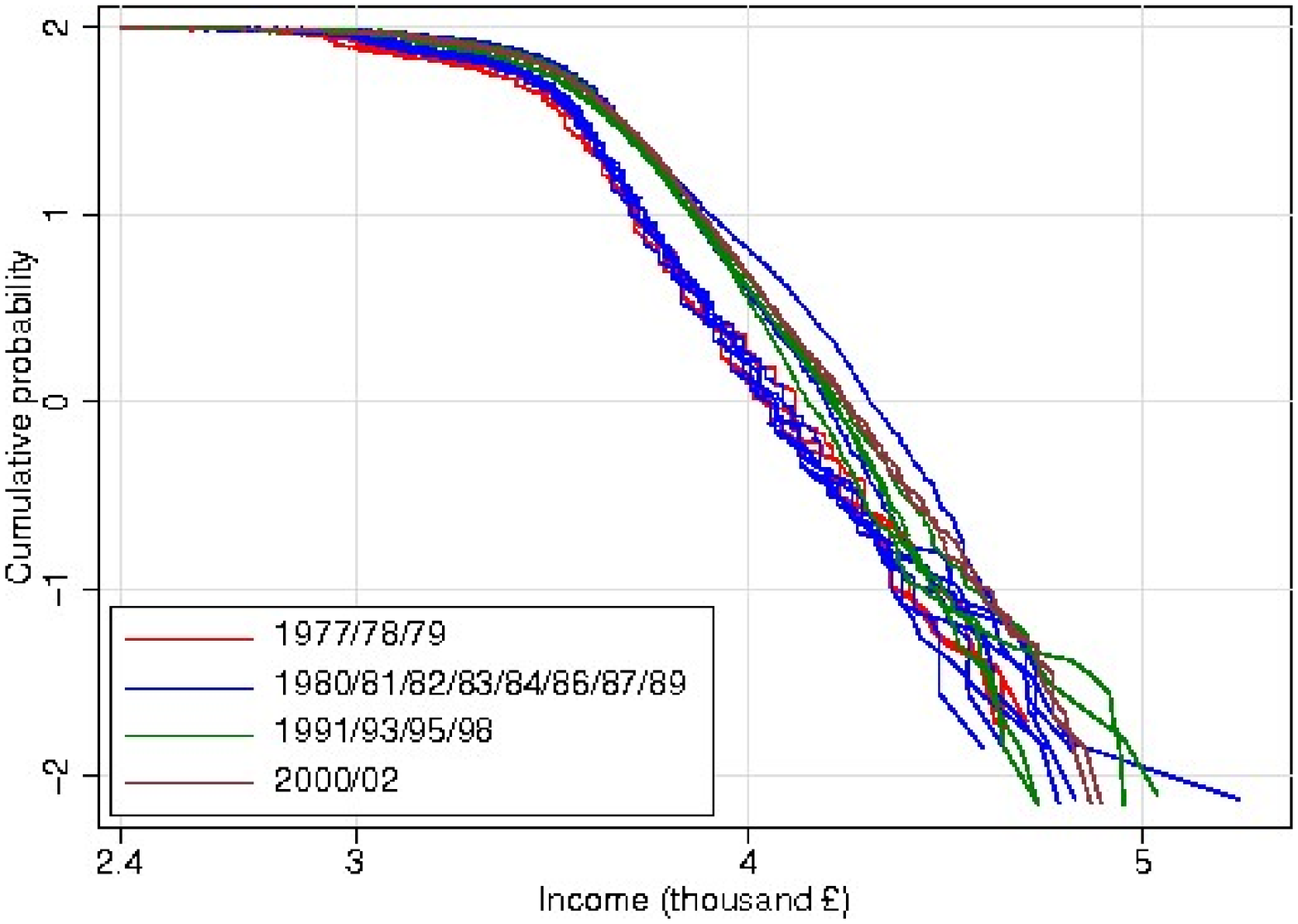}
\caption{Time development of the Italian personal income distribution over the years 1977--2002.}
\label{Time_Development_of_Income_Distribution}
\end{figure}

The distribution pattern of the personal income expressed as the lognormal with power law tails seems to hold all over our time span, as one can easily recognize from Fig. \ref{Time_Development_of_Income_Distribution},
which shows the shape of income distribution for all the years. The corresponding estimated parameters for the lognormal and Pareto distributions are given in Table \ref{Estimation_Results}.
\begin{table}[!t]
\centering
\caption{Estimated lognormal and Pareto distribution parameters for all the years (standard errors are in parentheses). Calculations are conducted using data and methods described in the text. For the lognormal distribution and the Pareto distribution also shown are the values of Gibrat index and $R^{2}$ index respectively.}
\label{Estimation_Results}
\begin{tabular}[width=\textwidth]{c c c c c c c}
\hline
Year & $\hat{\mu}$ & $\hat{\sigma}$ & $\hat{\beta}$ & $\hat{\alpha}$ & $\hat{x}_{0}$ & $R^{2}$\\[0.5ex]
\hline
1977 & 3.31 (0.005) & 0.34 (0.004) & 2.08 & 3.00 (0.008) & 10,876 & 0.9921\\
1978 & 3.33 (0.005) & 0.34 (0.004) & 2.09 & 3.01 (0.008) & 11,217 & 0.9933\\
1979 & 3.34 (0.005) & 0.34 (0.005) & 2.08 & 2.91 (0.009) & 11,740 & 0.9908\\
1980 & 3.36 (0.005) & 0.33 (0.005) & 2.15 & 3.06 (0.008) & 11,453 & 0.9915\\
1981 & 3.36 (0.005) & 0.32 (0.004) & 2.23 & 3.30 (0.008) & 10,284 & 0.9939\\
1982 & 3.38 (0.004) & 0.31 (0.005) & 2.27 & 3.08 (0.005) & 11,456 & 0.9952\\
1983 & 3.38 (0.004) & 0.30 (0.004) & 2.32 & 3.11 (0.006) & 11,147 & 0.9945\\
1984 & 3.39 (0.004) & 0.32 (0.005) & 2.24 & 3.05 (0.007) & 11,596 & 0.9937\\
1986 & 3.40 (0.004) & 0.29 (0.006) & 2.40 & 3.04 (0.005) & 11,597 & 0.9950\\
1987 & 3.49 (0.004) & 0.30 (0.004) & 2.38 & 2.09 (0.002) & 24,120 & 0.9993\\
1989 & 3.53 (0.003) & 0.26 (0.003) & 2.70 & 2.91 (0.002) & 15,788 & 0.9995\\
1991 & 3.52 (0.004) & 0.27 (0.004) & 2.58 & 3.45 (0.008) & 14,281 & 0.9988\\
1993 & 3.47 (0.004) & 0.33 (0.004) & 2.15 & 2.74 (0.002) & 16,625 & 0.9997\\
1995 & 3.46 (0.004) & 0.32 (0.003) & 2.19 & 2.72 (0.002) & 16,587 & 0.9996\\
1998 & 3.48 (0.004) & 0.34 (0.006) & 2.10 & 2.76 (0.002) & 17,141 & 0.9993\\
2000 & 3.50 (0.004) & 0.32 (0.004) & 2.20 & 2.76 (0.002) & 17,470 & 0.9994\\
2002 & 3.52 (0.004) & 0.31 (0.005) & 2.25 & 2.71 (0.002) & 17,664 & 0.9997\\[1ex]
\hline
\end{tabular}
\end{table}
The table also shows the values of Gibrat index and the OLS $R^{2}$. However, the power law slope and the curvature of the lognormal fit differ from each other. This fact means that the indexes specifying the distribution (Pareto and Gibrat indexes) differs from year to year. We therefore try to quantify the fluctuations of the shape of income
distribution in the next section.
\section{Time development of the distribution}
\label{sec:TimeDevelopmentOfTheDistribution}
\begin{figure}[!t]
\centering
\includegraphics[width=\textwidth]{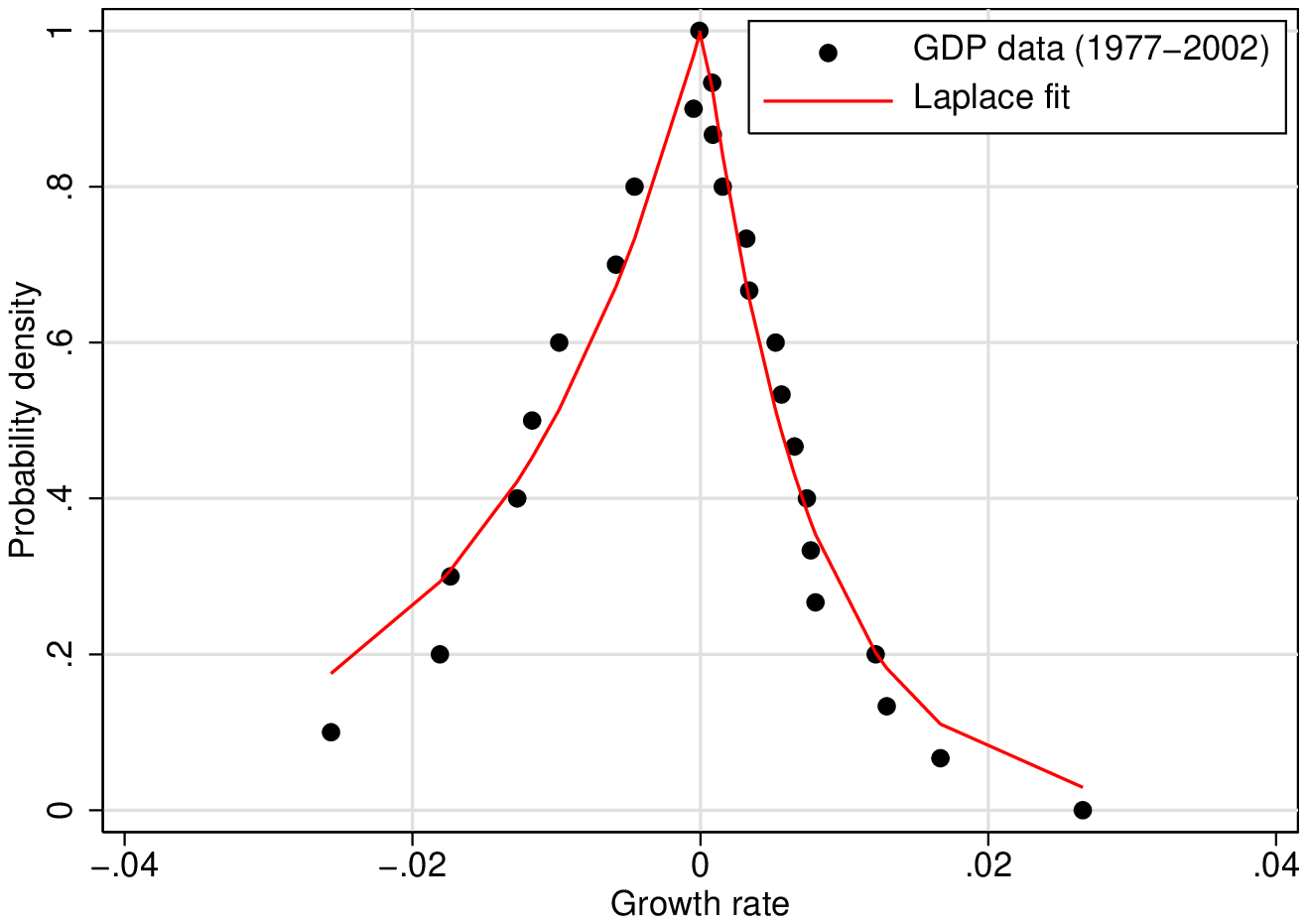}
\caption{Probability density function of Italian GDP annual growth rates for the period 1977--2002 together with the Laplace fit (red solid line). The data have been detrended by applying the Hodrick-Prescott filter.}
\label{GDP_Growth_Rate}
\end{figure}

\begin{figure}[!t]
\centering
\begin{minipage}[t]{0.5\textwidth}
\centering
\includegraphics[width=\textwidth]{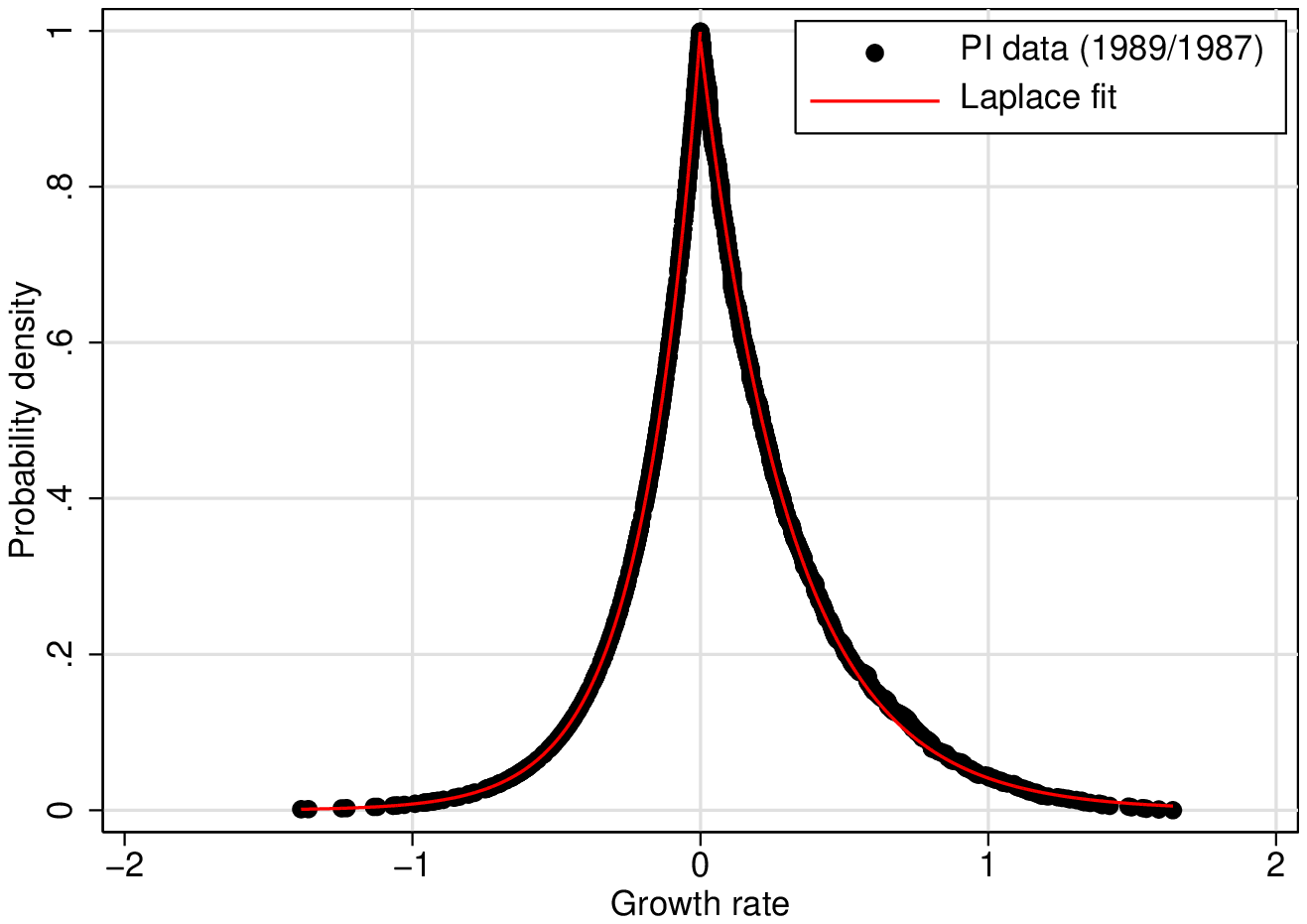}
\end{minipage}%
\centering
\begin{minipage}[t]{0.5\textwidth}
\centering
\includegraphics[width=\textwidth]{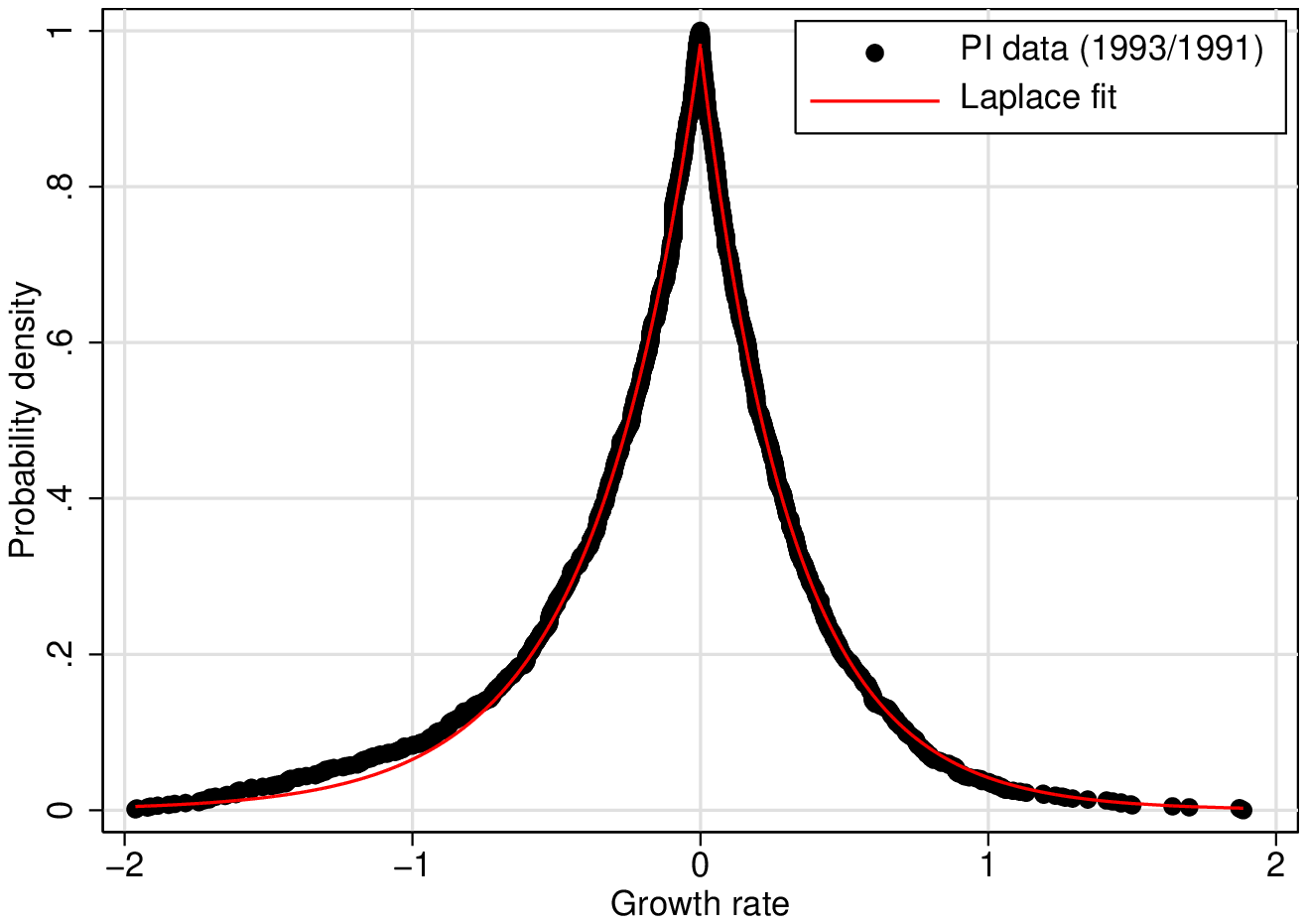}
\end{minipage}
\caption{Probability distribution of Italian PI growth rates $R_{\mathrm{PI}}\equiv log\left(PI_{t+i}/PI_{t}\right)$ for the years 1989/1987 and 1993/1991. The red solid line is the fit to the Laplace distribution. The data are taken from the panel section of the SHIW-HA, which covers the years 1987--2002.}
\label{PI_Growth_Rate_1989_1987_1993_1991}
\end{figure}

\begin{figure}[!t]
\centering
\includegraphics[width=\textwidth]{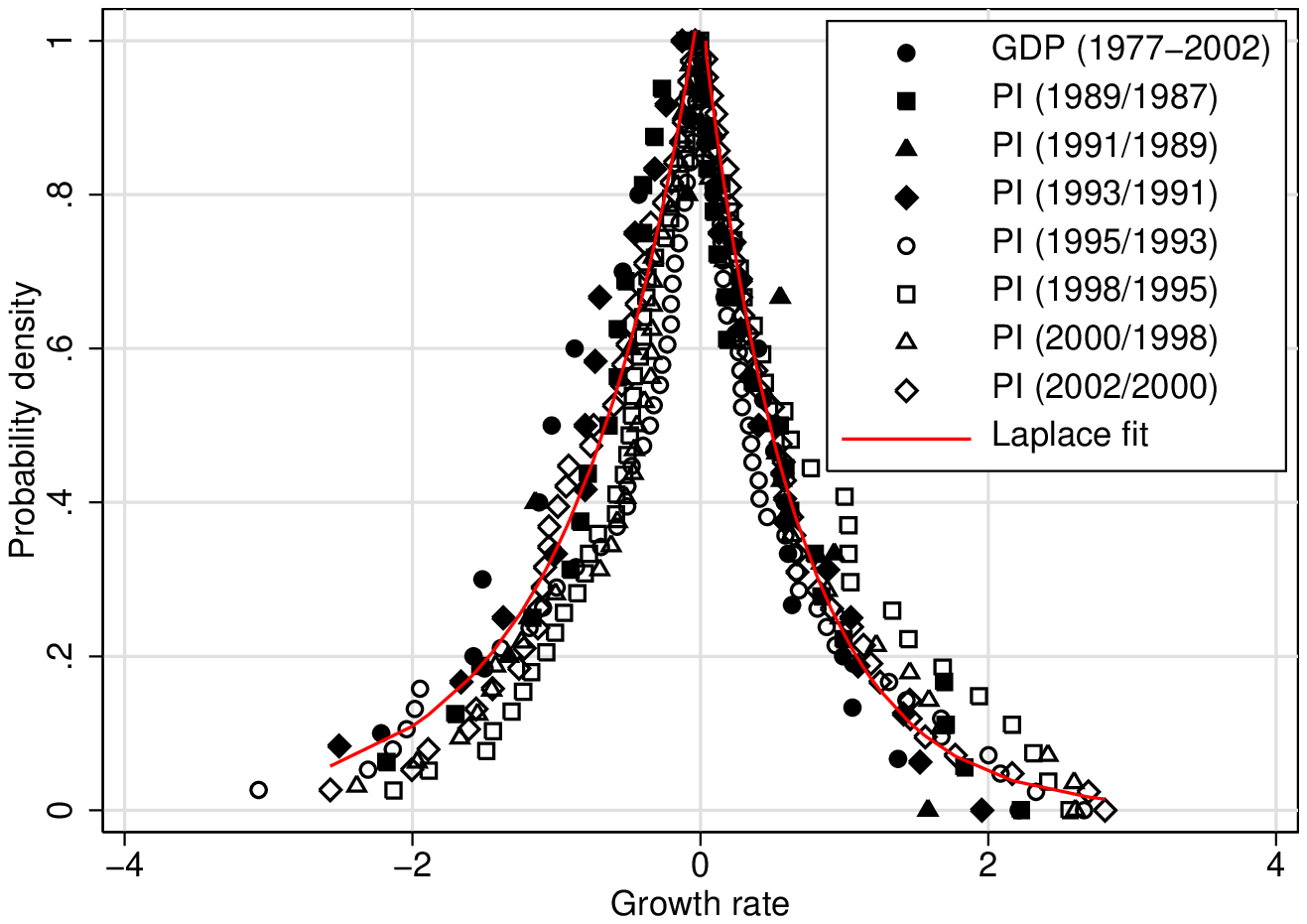}
\caption{Probability distribution of Italian GDP and PI growth rates. All data collapse onto a single curve representing the fit to the Laplace distribution (red solid line), showing that the distributions are well described by the same functional form.}
\label{GDP_PI_Growth_Rate}
\end{figure}

We start by considering the change of the distribution over time. As Fig. \ref{Time_Development_of_Income_Distribution}
shows, the distribution shifts over the years covered by our dataset. Macroeconomics argues that the origin of the change consists in the growth of the Gross Domestic Product (GDP). To confirm this hypothesis we study the fluctuations in the growth rates of GDP and Personal Income (PI), and try to show that similar mechanisms may be responsible for the observed growth dynamics of both country and individuals. The distribution of GDP annual growth rates is shown in Fig. \ref{GDP_Growth_Rate}.
We calculate it using the data from the OECD Statistical Compendium \cite{OECD}, and expressing the rates in terms of
their logarithm, $R_{\mathrm{GDP}}\equiv log\left(GDP_{t+1}/GDP_{t}\right)$, where $GDP_{t}$ and $GDP_{t+1}$ are the GDP of the country at the years $t$ and $t+1$ respectively. Data are reported in 1976 prices; moreover, to improve comparison of the values over the years we detrend them by applying the Hodrick-Prescott filter. By means of a non-linear algorithm, we find that the probability density function of annual growth rates is well fitted by a Laplace distribution (the red solid line in the figure), which is expressed as:
\begin{equation}
p\left(x\right)=\frac{1}{\sigma\sqrt{2}}exp\left(-\frac{\left|x-\mu\right|}{\sigma}\right)
\end{equation}
with $-\infty<x<+\infty$, and where $\mu$ and $\sigma$ are the mean value and the standard deviation. This result seems to be in agreement with the growth dynamics of PI, as shown in Fig. \ref{PI_Growth_Rate_1989_1987_1993_1991}
for two randomly selected distributions. We calculate them using the panel section of the SHIW-HA, which covers the period 1987--2002. As one can easily recognize, the same functional form describing the probability distribution of GDP annual growth rates is also valid in the case of PI growth rates. These findings lead us to check the possibility
that the growth rates of both GDP and PI are drawn from the same distribution. To this end, we perform a two-sample Kolmogorov-Smirnov test and check the null hypothesis that both GDP and PI growth rate data are samples from the same distribution. Before applying this test, to consider almost the same number of data points relating to units with different sizes we draw a 2\% random samples of the data we have for individuals, and normalize them together with the data for GDP annual growth rate using the transformations $\left(R_{\mathrm{PI}}-\overline{R_{\mathrm{PI}}}\right)/\sigma_{\mathrm{PI}}$ and $\left(R_{\mathrm{GDP}}-\overline{R_{\mathrm{GDP}}}\right)/\sigma_{\mathrm{GDP}}$. As shown in Table \ref{Kolmogorov_Smirnov_Test},
\begin{table}[!t]
\centering
\caption{Estimated Kolmogorov-Smirnov test {\em p\/}-values for both GDP and PI growth rate data. The null hypothesis that the two distributions are the same at the 5\% marginal significance level is not rejected in all the cases.}
\label{Kolmogorov_Smirnov_Test}
\begin{tabular}[width=\textwidth]{c c c c c c c c}
\hline
Growth rate & $R_{89/87}$ & $R_{91/89}$ & $R_{93/91}$ & $R_{95/93}$ & $R_{98/95}$ & $R_{00/98}$ & $R_{02/00}$\\[0.5ex]
\hline
$R_{\mathrm{GDP}}$ & 0.872 &	0.919	& 0.998	& 0.696	& 0.337	& 0.480	& 0.955\\
$R_{89/87}$ & & 0.998	& 0.984	& 0.431	& 0.689	& 0.860	& 0.840\\
$R_{91/89}$ & & & 0.970	& 0.979	& 0.995	& 0.994	& 0.997\\
$R_{93/91}$ & & & & 0.839	& 0.459	& 0.750	& 1.000\\
$R_{95/93}$ & & & & & 0.172	& 0.459	& 0.560\\
$R_{98/95}$ & & & & & & 0.703	& 0.378\\
$R_{00/98}$ & & & & & & & 0.658\\[1ex]
\hline
\end{tabular}
\end{table}
which reports the {\em p\/}-values for all the cases we studied, the null hypothesis of equality of the two distributions can not be rejected at the usual 5\% marginal significance level. Therefore, the data are consistent with the assumption that a common empirical law might describe the growth dynamics of both GDP and PI, as shown in Fig. \ref{GDP_PI_Growth_Rate},
where all the curves for both GDP and PI growth rate normalized data almost collapse onto the red solid line representing the non-linear Laplace fit.\footnote{See \cite{Lee_Amaral_Canning_Meyer_Stanley} for similar findings about GDP and company growth rates.}
\par
\begin{figure}[!t]
\centering
\begin{minipage}[t]{0.5\textwidth}
\centering
\includegraphics[width=\textwidth]{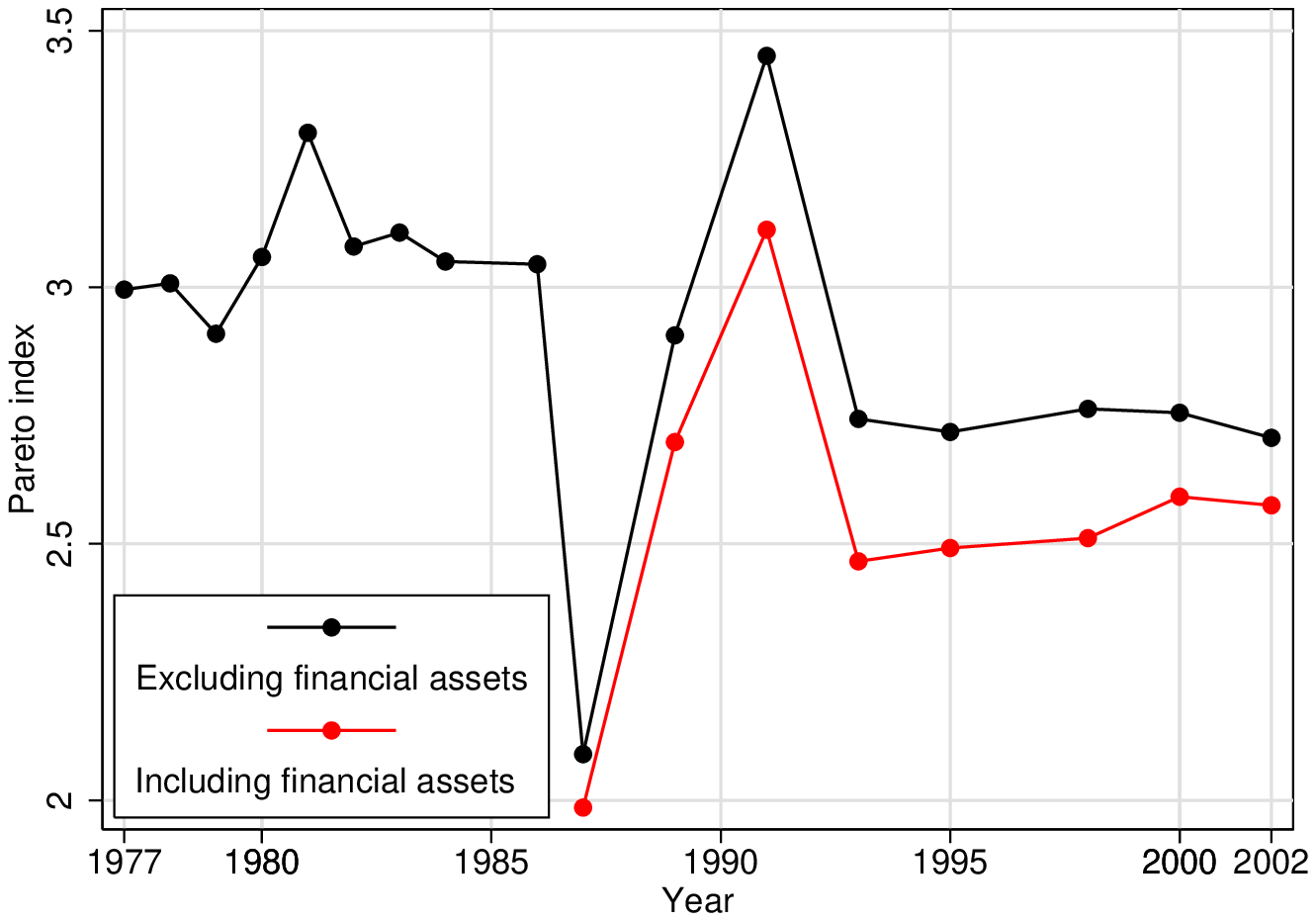}
\end{minipage}%
\centering
\begin{minipage}[t]{0.5\textwidth}
\centering
\includegraphics[width=\textwidth]{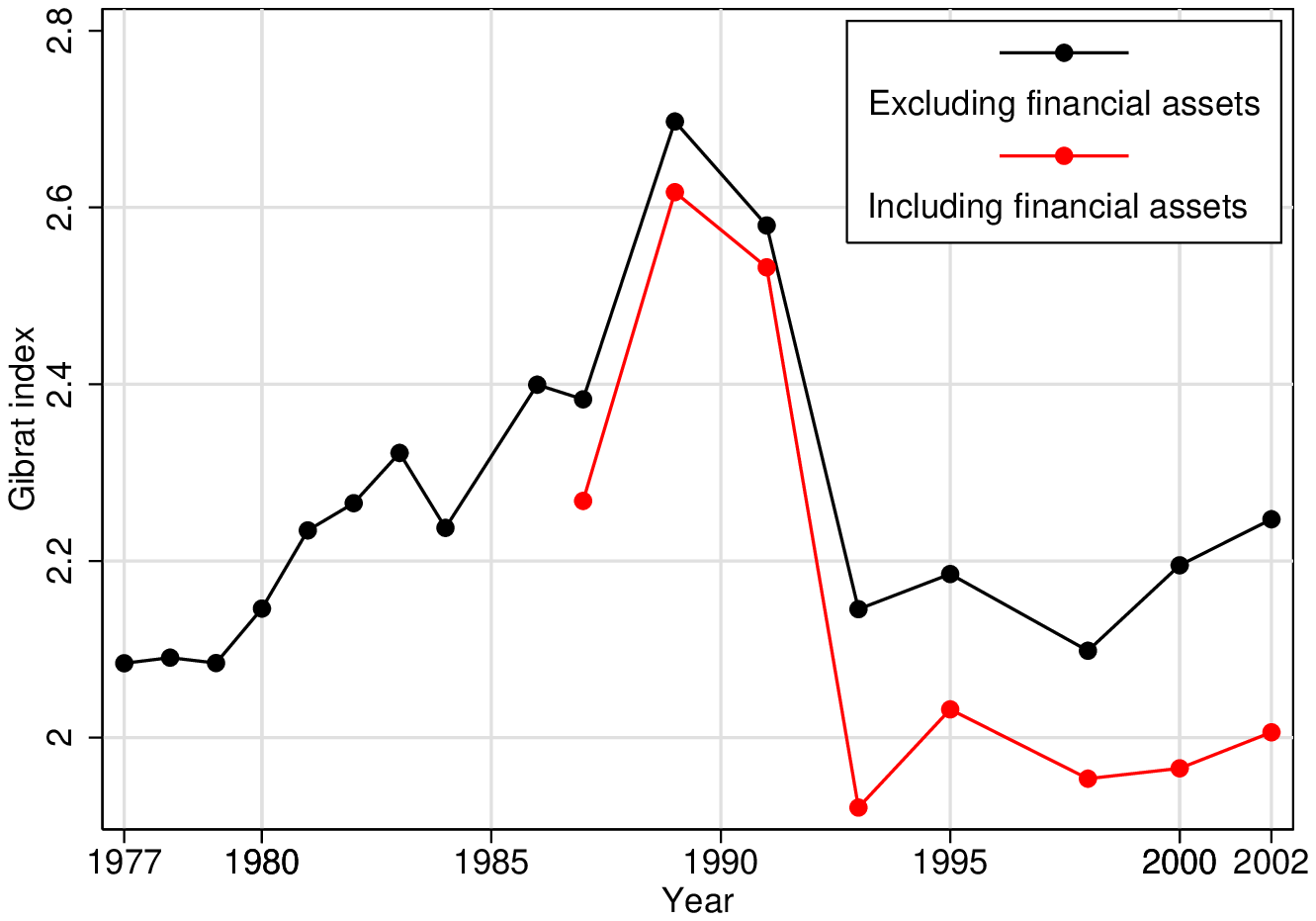}
\end{minipage}
\caption{The temporal change of Pareto index (left panel) and Gibrat index (right panel) over the years 1977--2002. The black connected solid line is the time series of the two indexes obtained by excluding income from financial assets, while the red connected solid line refers to the yearly estimates obtained by the inclusion of the above-stated income.}
\label{Fluctuations_of_Pareto_and_Gibrat_Indexes}
\end{figure}
\begin{figure}[!t]
\centering
\includegraphics[width=\textwidth]{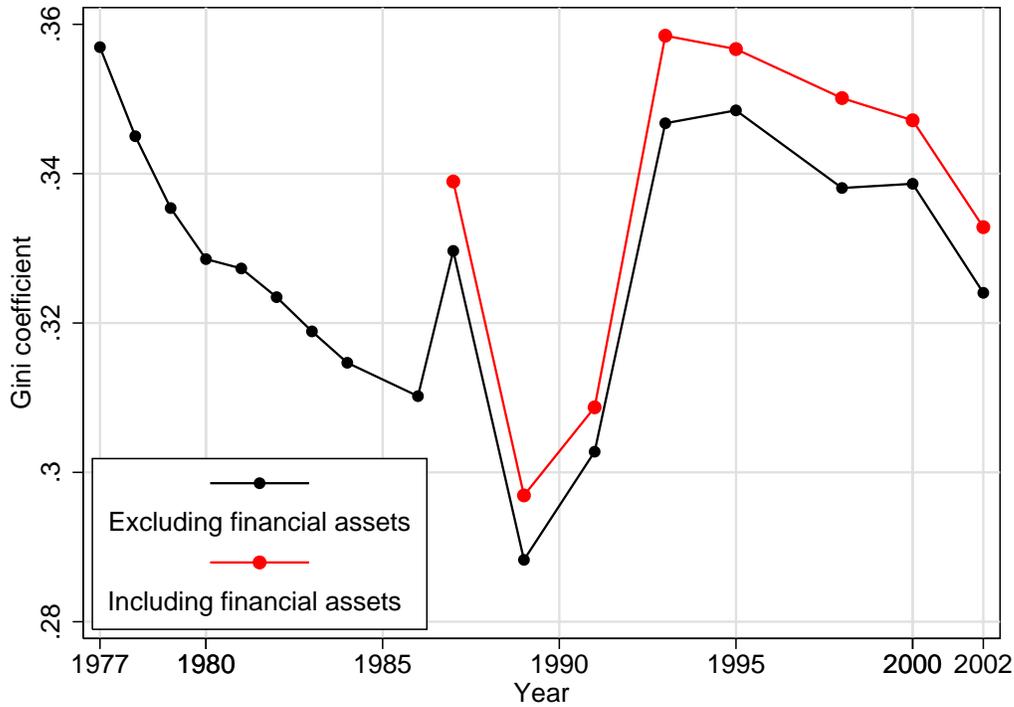}
\caption{Gini coefficient for Italian personal income during the period 1977--2002. The black connected solid line excludes income from financial assets, while the red connected solid line includes it.}
\label{Gini_Coefficient}
\end{figure}
\begin{figure}[!t]
\centering
\begin{minipage}[t]{0.5\textwidth}
\centering
\includegraphics[width=\textwidth]{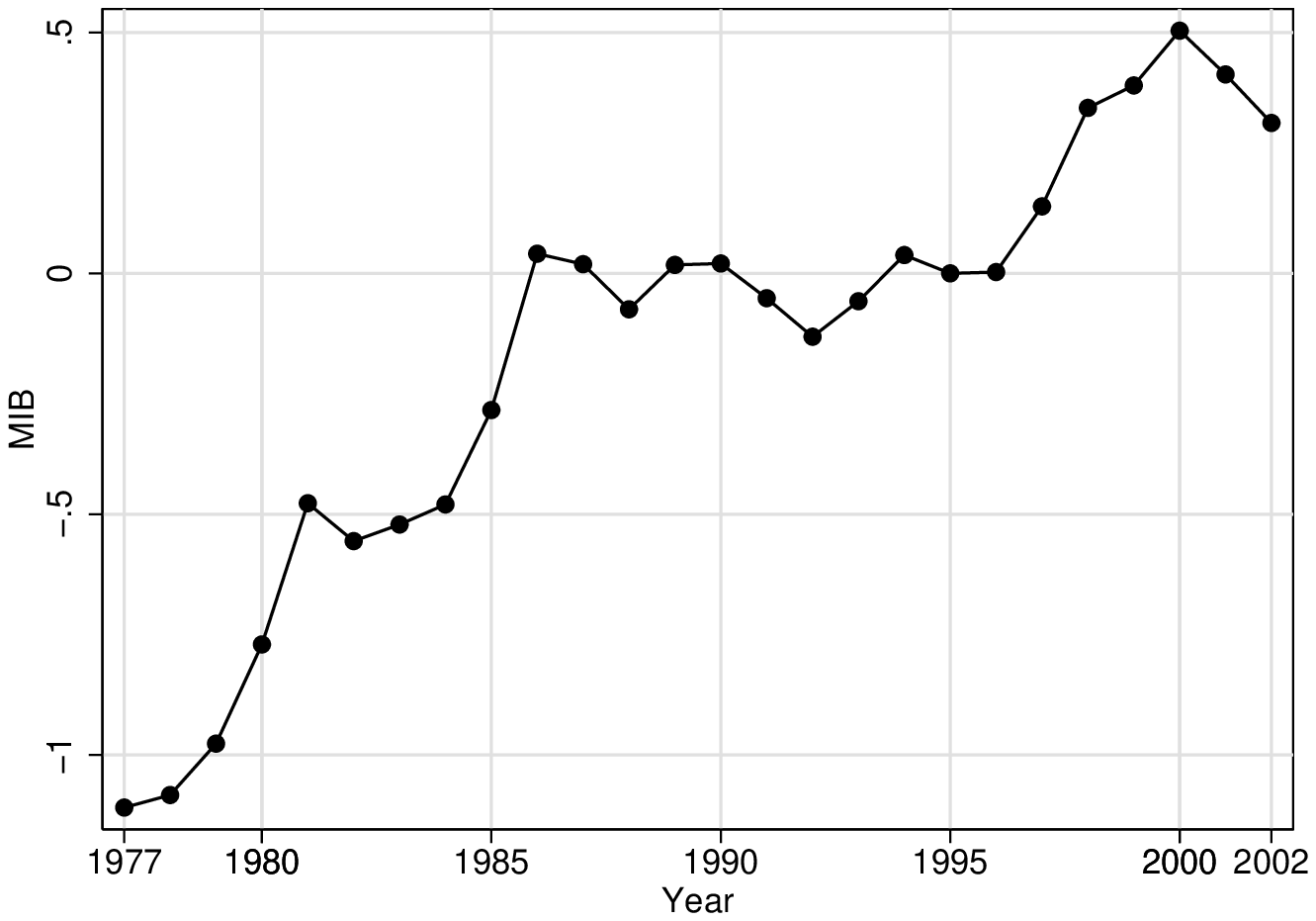}
\end{minipage}%
\centering
\begin{minipage}[t]{0.5\textwidth}
\centering
\includegraphics[width=\textwidth]{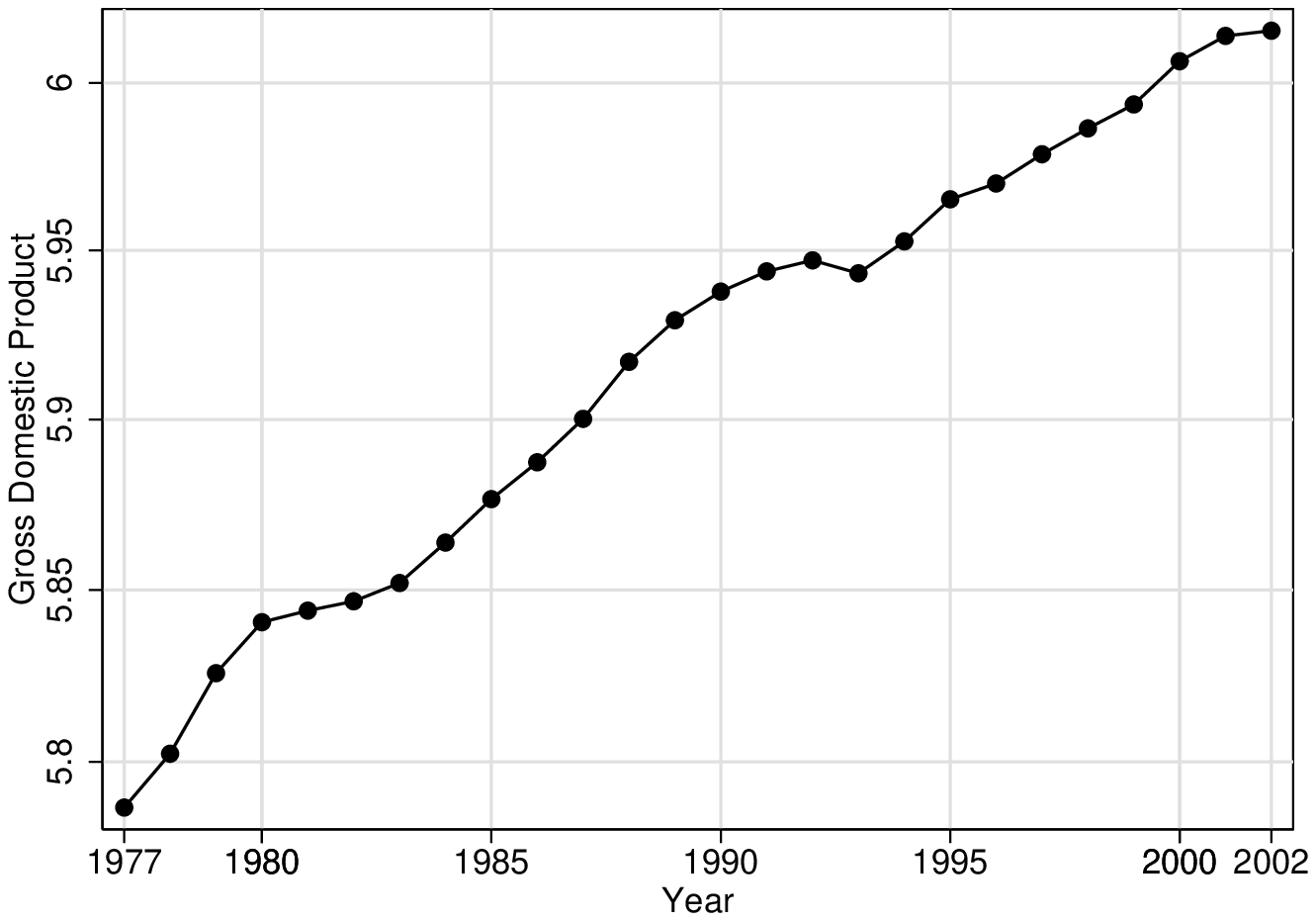}
\end{minipage}
\caption{Temporal change of Italian Stock Exchange MIB Index (left panel) and Italian GDP (right panel) during the period 1977--2002. Shown are the logarithmic values of the variables. The GDP is with the unit of million EUR at 1995 prices. The data source is \cite{OECD}.}
\label{Fluctuations_of_Historical_MIB_Index_and_GDP}
\end{figure}
\begin{figure}[!t]
\centering
\begin{minipage}[t]{0.5\textwidth}
\centering
\includegraphics[width=\textwidth]{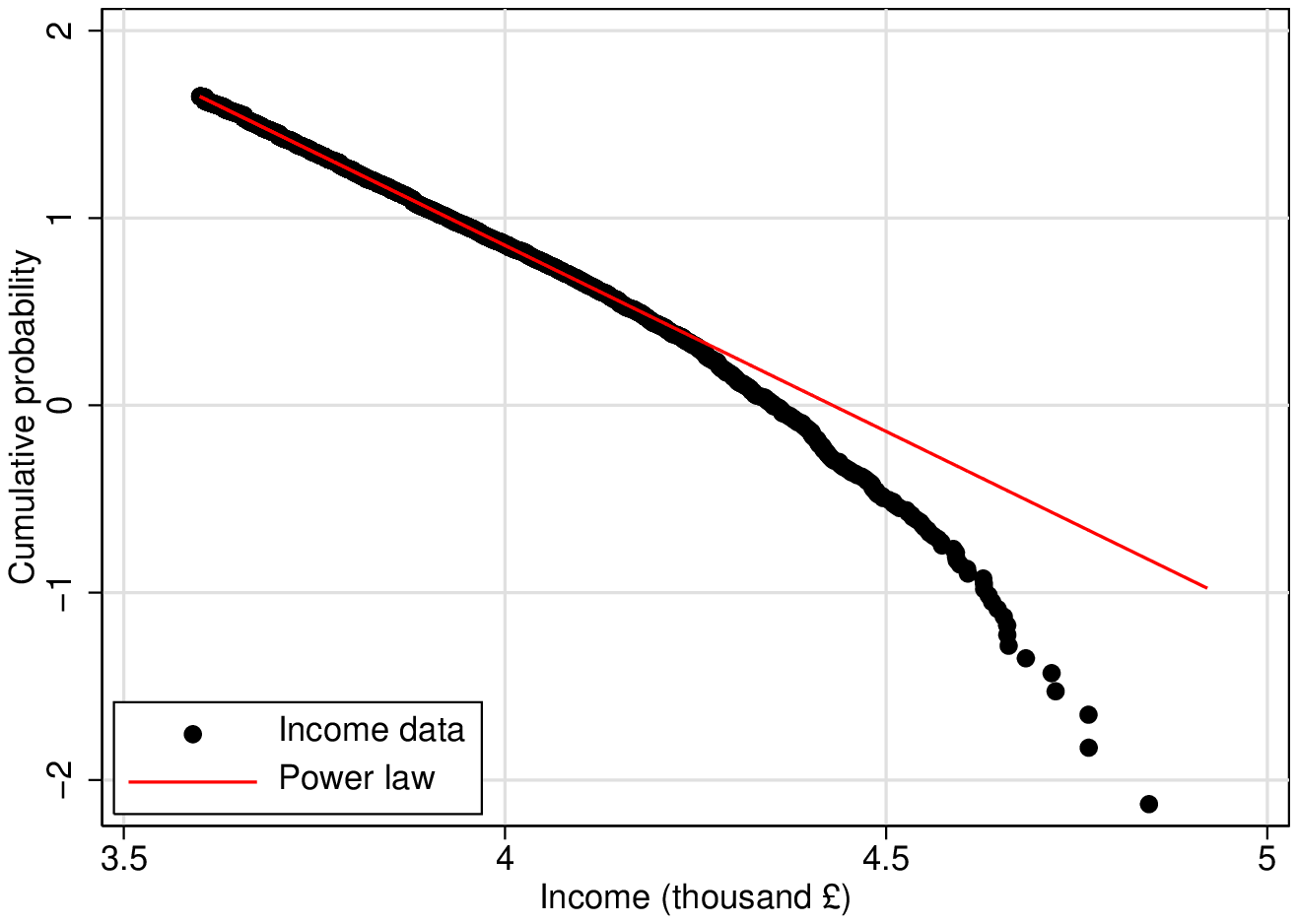}
\end{minipage}%
\centering
\begin{minipage}[t]{0.5\textwidth}
\centering
\includegraphics[width=\textwidth]{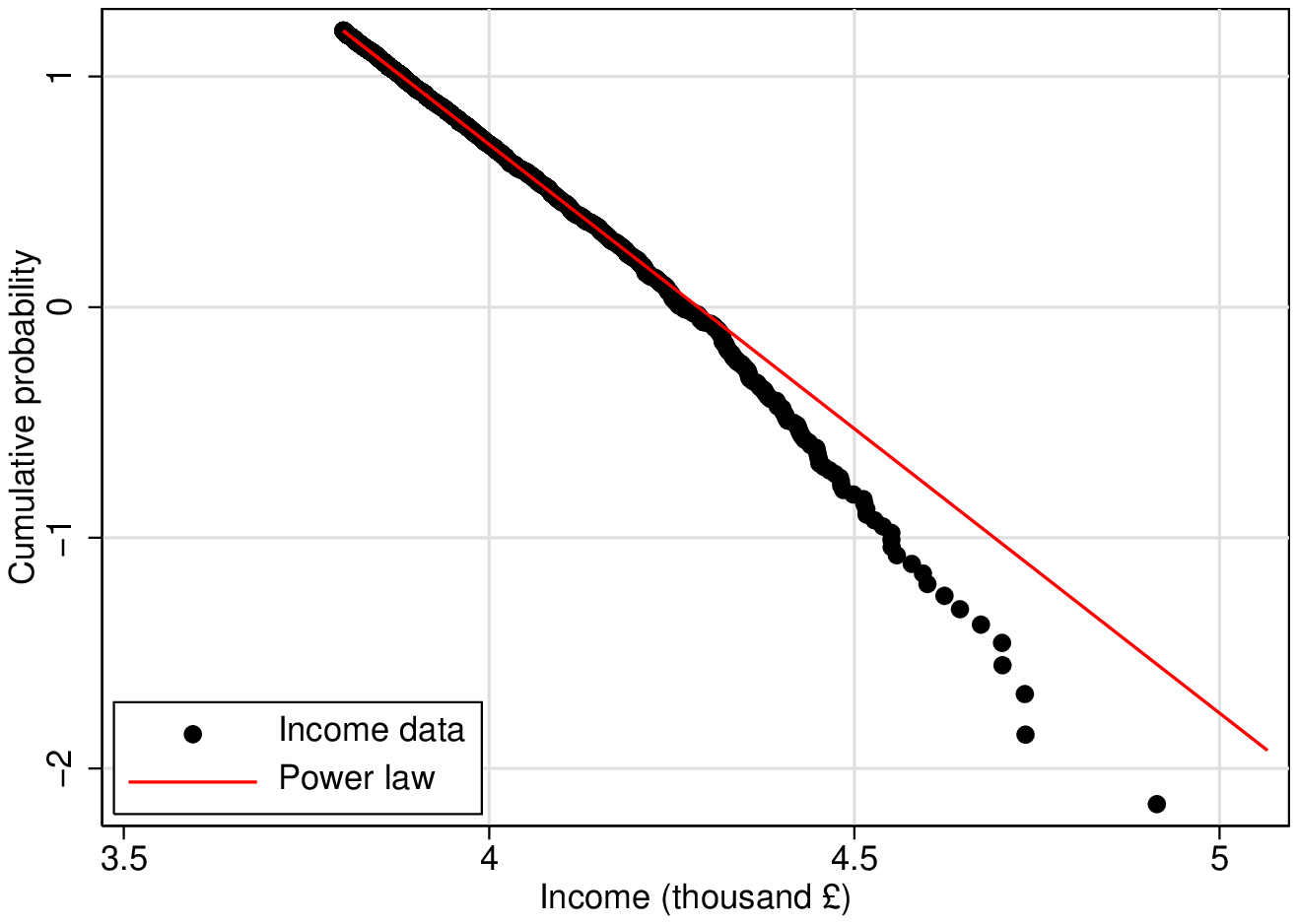}
\end{minipage}
\caption{The power law region in 1987 and 1993. The red solid line is the fit to the power law distribution. The Pareto index for 1987 data is $\hat{\alpha}=2.09$ $(0.002)$, while for 1993 data we have $\hat{\alpha}=2.74$ $(0.002)$. These estimates were obtained by least-square-fit excluding about top and bottom 1.2\% for 1987 data, and about top 1.4\% and bottom 1.3\% for 1993 data. OLS $R^{2}$ values are 0.9993 and 0.9997 respectively. As one can note, a large deviation from Pareto law is seen in both the years.}
\label{Power_Law_1987_1993}
\end{figure}

We next turn on the fluctuations of the indexes specifying the income distribution, \ie{} the Pareto and Gibrat
indexes, whose yearly estimates are reported in Fig. \ref{Fluctuations_of_Pareto_and_Gibrat_Indexes}.
The left panel shows the fluctuations of Pareto index over the years 1977--2002. The black connected solid line is the time series obtained by excluding income from financial assets, while the red connected solid line refers to the yearly estimates obtained by the inclusion of the above-stated income, which was regularly recorded only since 1987 (see note 2). The course of the two series is similar, with the more complete definition of income showing a greater inequality because of the strongly concentrated distribution of returns on capital. The same can be said for the time series of Gibrat index (right panel). Although the frequency of data (initially annual and then biennial from 1987) makes it difficult to establish a link with the business cycle, it seems possible to find a (negative) relationship between the above-stated indexes and the fluctuations of economic activity. For example, Italy experienced a period of economic growth until the late `80s, but with alternating phases of the internal business cycle: of slowdown of production
up to the 1983 stagnation; of recovery in 1984; again of slowdown in 1986. As one can recognize from the figure, the values of Pareto and Gibrat indexes, inferred from the numerical fitting, tend to decrease in the periods of economic expansion (concentration goes up) and increase during the recessions (income is more evenly distributed). The time pattern of inequality is shown in Fig. \ref{Gini_Coefficient},
which reports the temporal change of Gini coefficient for the considered years.\footnote{Unlike the Pareto and Gibrat indexes, which provide two different measures corresponding to the tail and the rest, the Gini coefficient is a measure of (in)equality of the income for the overall distribution taking values from zero (completely equal) to one (completely inequal).} In Italy the level of inequality decreased significantly during the `80s and rised in the early
`90s; it was substantially stable in the following years. In particular, a sharp rise of Gini coefficient (\ie{}, of inequality) is encountered in 1987 and 1993, corresponding to a sharp decline of Pareto index in the former case and of both Pareto and Gibrat indexes in the latter case. We consider that the decline of Pareto exponent
in 1987 corresponds with the peak of the speculative `bubble' begun in the early `80s, and the rebounce of the index follows its burst on October 19, when the Dow Jones index lost more than 20\% of its value dragging into disaster the other world markets. This assumption seems confirmed by the movement of asset price in the Italian Stock Exchange
(see Fig. \ref{Fluctuations_of_Historical_MIB_Index_and_GDP},
left panel).\footnote{See \cite{Souma} for a study of the correlation between Pareto index and asset price in Japan.} As regards the sharp decline of both indexes in 1993, the level and growth of personal income (especially in the middle-upper income range) were notably influenced by the bad results of the real economy in that year, following the September 1992 lira exchange rate crisis. The effects of recession (visible in Fig. 9, right panel) produced a leftwards shift of the distribution and widened its range; this, combined with a concentration of individuals towards middle income range, induced an increase in inequality.\footnote{In particular, in this year there was a significant reduction of the number of self-employees, whose incomes are much more dependent from the business cycle. See 
\cite{Bank_of_Italy_2} for further details on this issue.} It would be expected that these facts cause the invalidity of
Pareto law for high incomes. This was well the case of Italian economy during the mentioned years. Fig. \ref{Power_Law_1987_1993}
shows the power law region in 1987 and 1993. Compared to other years, one can observe that the data can not be fitted by the Pareto law in the entire range of high-income.
\section{Concluding remarks}
\label{sec:ConcludingRemarks}
In this paper we find that the Italian personal income microdata are consistent with a Pareto-power law behaviour in the high-income range, and with a two-parameter lognormal pattern in the low-middle income region.
\par
The numerical fitting over the time span covered by our dataset show a shift of the distribution, which is claimed to be a consequence of the growth of the country. This assumption is confirmed by testing the hypothesis that the
growth dynamics of both gross domestic product of the country and personal income of individuals is the same; the two-sample Kolmogorov-Smirnov test we perform on this subject lead us to accept the null hypothesis that the
growth rates of both the quantities are samples from the same probability distribution in all the cases we studied, pointing to the existence of correlation between them.
\par
Moreover, by calculating the yearly estimates of Pareto and Gibrat indexes, we quantify the fluctuations of the shape of the distribution over time by establishing some links with the business cycle phases which Italian economy
experienced over the years of our concern. We find that there exists a negative relationship between the above-stated indexes and the fluctuations of economic activity at least until the late `80s. In particular, we show that in two circumstances (the 1987 burst of the asset-inflation `bubble' begun in the early `80s and the 1993 recession year) the data can not be fitted by a power law in the entire high-income range, causing breakdown of Pareto law.
\begin{ack}
The authors would like to thank Corrado Di Guilmi and Yoshi Fujiwara for helpful comments and suggestions.
\end{ack}

\end{document}